\documentclass[aps,prl, twocolumn, amsmath,amssymb,superscriptaddress]{revtex4}

\usepackage{graphicx}
\usepackage{dcolumn}
\usepackage{bm}
\usepackage{color}
\usepackage{pdfsync}
\usepackage{hyperref}

\begin{document}


\title{Tuned transition from quantum to classical for macroscopic quantum states}

\author{A.~Fedorov}
\email{fedoroar@phys.ethz.ch}
\altaffiliation{Present address: Department of Physics, ETH Zurich, CH-8093, Zurich, Switzerland}
\affiliation{Kavli Institute of Nanoscience, Delft University of Technology, PO Box 5046, 2600 GA Delft, The Netherlands}
\author{P.~Macha}
\affiliation{Institute of Photonic Technology, P.O. Box 100239, D-07702 Jena, Germany}
\affiliation{Kavli Institute of Nanoscience, Delft University of Technology, PO Box 5046, 2600 GA Delft, The Netherlands}
\author{A.~K.~Feofanov}
\affiliation{Physikalisches Institut and DFG Center for Functional Nanostructures (CFN) Karlsruhe Institute of Technology, Wolfgang-Gaede-Str. 1, D-76131 Karlsruhe, Germany}
\affiliation{Kavli Institute of Nanoscience, Delft University of Technology, PO Box 5046, 2600 GA Delft, The Netherlands}
\author{C. J. P. M.~Harmans}
\affiliation{Kavli Institute of Nanoscience, Delft University of Technology, PO Box 5046, 2600 GA Delft, The Netherlands}
\author{J. E.~Mooij}
\affiliation{Kavli Institute of Nanoscience, Delft University of Technology, PO Box 5046, 2600 GA Delft, The Netherlands}

\begin{abstract}
The boundary between the classical and quantum worlds has been intensely studied. It remains fascinating to explore how far the quantum concept can reach with use of specially fabricated elements.  Here we employ a tunable flux qubit with basis states having persistent currents of 1$~\mu$A  carried by a billion electrons. By tuning the tunnel barrier between these states we see a cross-over from quantum to classical. Released from non-equilibrium, the system exhibits spontaneous coherent oscillations. For high barriers the lifetime of the states increases dramatically while the tunneling period approaches the phase coherence time and the classical regime is reached.
\end{abstract}
\maketitle
The quantum nature of quarks and atoms is as solidly established as the relevance of Newtonian mechanics for marbles and soccer balls. The boundary between the two worlds has been studied theoretically~\cite{Leggett review}, but only a few experiments are available so far. It has been demonstrated that objects containing many atoms, such as large molecules \cite{Molecules}, magnetic particles \cite{Magnets} or fabricated superconducting circuits~\cite{Cooper-pair box,Sacley L-G} can behave like single quantum particles. In this Letter we performed  an experiment on a superconducting flux qubit, which is the 'classical' example of a macroscopic object that can be made to behave as a quantum particle. It is characterized by two states with opposite macroscopic currents in a loop. We were able to control the tunnel barrier between these states over a very wide range. We tuned qubit energy levels below the barrier and the same time effectively cool the sample to near zero temperature. This allowed us to study the qubit behavior when we go from the range of low barriers and strong quantum tunnelling to the regime where
quantum tunnelling gradually disappears as the barrier is increased. In particular, we manage to observe the natural quantum oscillations manifested in the tunnelling of the long-living macroscopic magnetic moments. At very high barriers we see the how these oscillations fade away as the barrier increased.


\begin{figure}[t!]
\includegraphics[width=75mm]{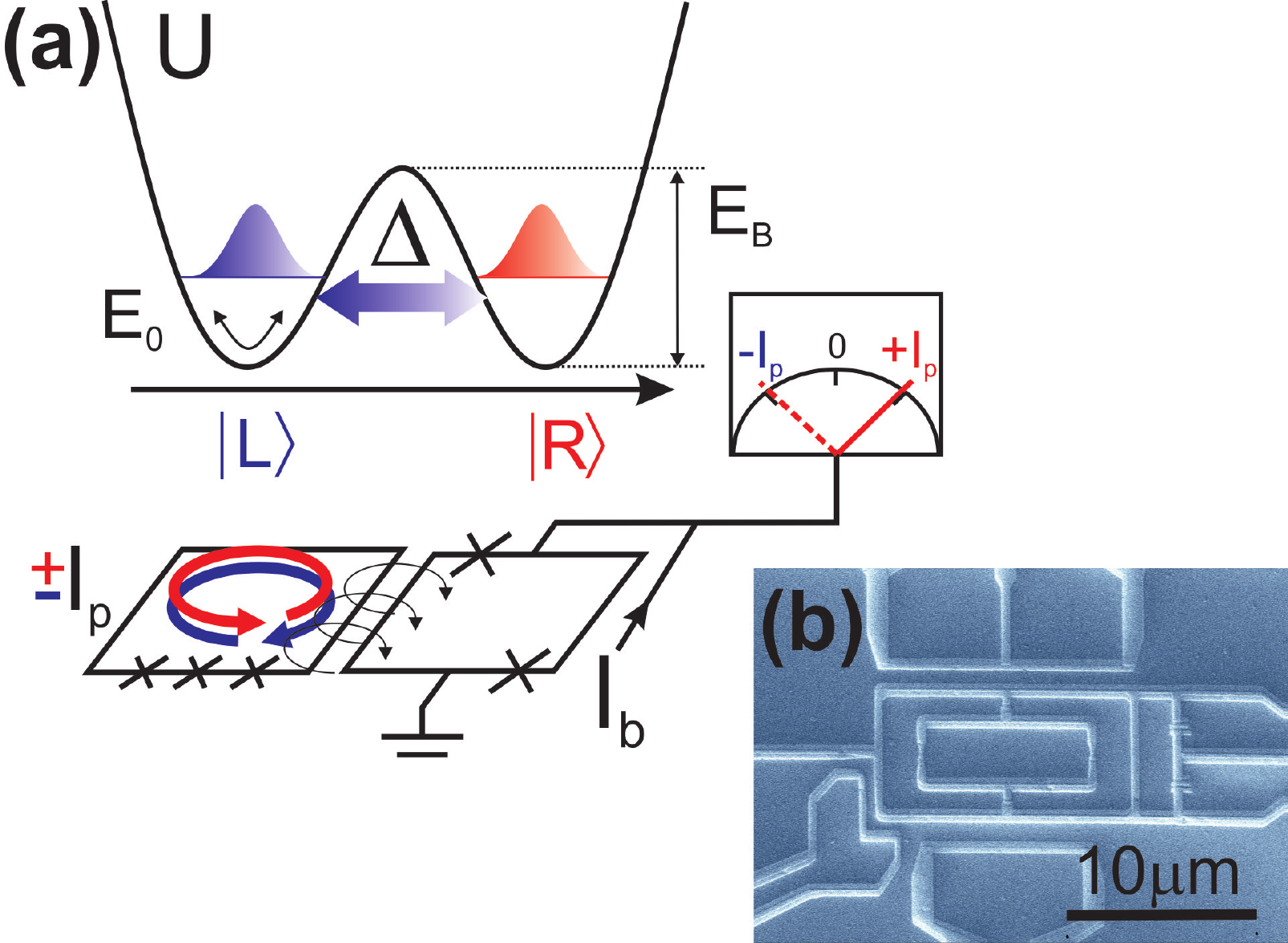}
\caption{\label{double well} Experiment to test macroscopic quantum coherence: (a) Potential energy of the flux qubit.  The barrier height $E_B$ is large compared to the zero-point energy $E_0$ resulting in the strong localization of the well ground states $|L\rangle$ and $|R\rangle$.  These states are connected to macroscopically distinguishable magnetic moments induced by the persistent currents in the flux qubit loop. The tunnel coupling $\Delta$ between the current states is exponentially dependent on $E_B$. The macroscopic quantum tunnelling $|L\rangle\leftrightarrow|R\rangle$ can be observed by a detector (DC-SQUID) sensitive to a change of the magnetic flux in the qubit loop. (b) Scanning electron micrograph of the flux qubit.}
\end{figure}
\begin{figure}[t!]
\includegraphics[width=75mm]{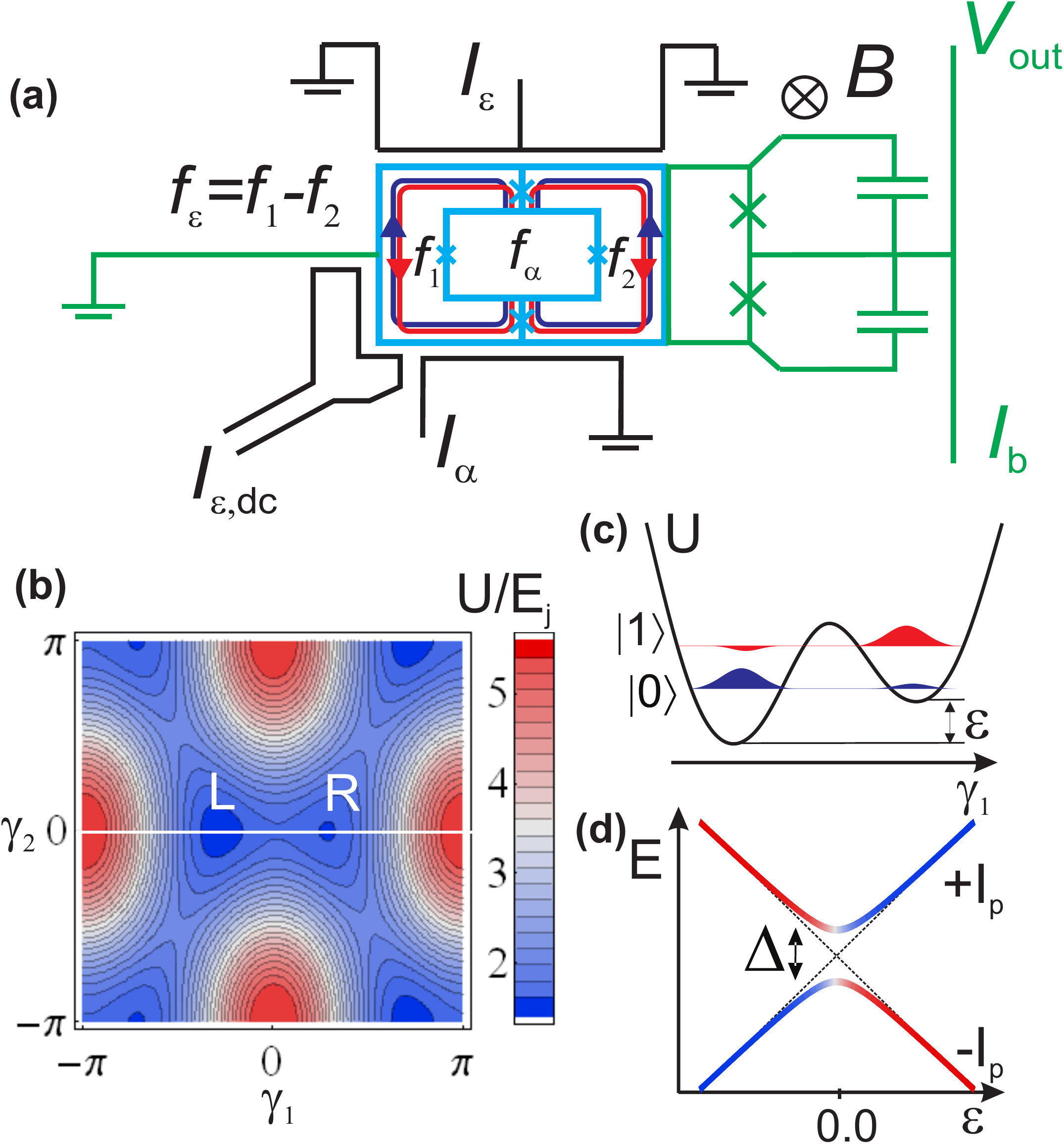}
\caption{\label{scheme} Tunable flux qubit: (a) Schematics. The qubit (blue) is formed by three Josephson junctions one of which is a tunable double junction. The dark blue and red arrowed lines show the persistent currents connected to the states $|L\rangle$ and $|R\rangle$. The qubit state can be controlled by the bias lines $I_\varepsilon$, $I_{\varepsilon,dc}$, $I_\alpha$ (black) and measured by  the DC SQUID (green).
(b) Potential energy (in units of the Josephson energy of the regular junctions) as a function of the two independent phase differences $\gamma_1$ and $\gamma_2$. (c) Sketch of the cross section of the potential energy along the white line connecting left and right wells through the saddle point (see (b). The energy eigenstates $|0\rangle$ and $|1\rangle$ are superpositions of the persistent current states $|L\rangle$ and $|R\rangle$. (d) Energy diagram of the qubit vs. magnetic bias $\epsilon$.}
\end{figure}


The flux qubit has a potential energy which consists of two degenerate wells [Fig.~\ref{double well}(a)] separated by a barrier $E_B$.
Each well is connected with a macroscopic magnetic flux, with a sign ($+/-$) depending on being in the left or the right well, which can be  detected on demand by a measurement apparatus. The  zero-point energy $E_0$ of the qubit in each well can be made smaller than $E_B$~\cite{Caspar,Fluxonium}.
Consequently, the barrier between the wells becomes classically impenetrable, and at low temperature the magnetic moment of the qubit can be flipped only via the quantum tunnelling process. This process is represented in Fig.~\ref{double well}(a) by the tunnelling coupling $\Delta$, which depends exponentially on the barrier height $E_B$.
Thus, the flux qubit is particularly suited for testing quantum-to-classical crossover, as it integrates two seemingly contradictory features of the classical and quantum world: the macroscopic character of the quantum states with the fundamentally non-classical  quantum tunnelling.
Other attempts to study the quantum-to-classical transition on a single quantum system were based on increasing the effect of the noise on a system with fixed quantum interactions \cite{Wallraff}.  To probe the quantum nature of the qubit we use the `real-time' experiment proposed by A.~Leggett~\cite{Leggett review}: prepare the system in one well, let the system evolve for a  time $t$ and measure the magnetic flux with the detector. The resulting quantum mechanical probability to find the system in the initial well equals $P(t)=\left(1+\cos(2\pi \Delta t)\right)/2$. Observation of  the magnetic flux oscillations should directly prove the `quantumness'. There were two earlier experiments which used similar measurement protocols for superconducting qubits in the charge~\cite{Cooper-pair box} and phase~\cite{Karlsruhe} regimes. Both experiments were performed only in the pure quantum regime for $h\nu_{\rm qb}\gg k_B T_{\rm bath}$ where $\nu_{\rm qb}$ is the qubit energy splitting and $T_{\rm bath}$ is the temperature of the qubit environment. In the former experiment the observed oscillations were attributed to tunnelling of a single Cooper-pair. In the latter one the quantum oscillations were observed between states not connected to a macroscopic variable. Furthermore, in a flux qubit experiment a qualitatively different behavior for different tunnel barriers was attributed to the quantum and classical regimes~\cite{Grajcar}.
Up to now no real time observation of the macroscopic flux tunnelling has been reported.

Our flux qubit consists of three junctions symmetrically attached to a trap loop as shown in Fig.~\ref{scheme}(a). The central junction is made tunable by replacing it by two junctions in parallel, thus providing control over $E_B$ and so $\Delta$.
The trap loop is employed to capture a fluxoid (or 2$\pi$-phase-winding number)~\cite{fluxon trap}, establishing a $\pi$ phase drop over the qubit junctions. If one fluxoid is trapped and the difference in flux in the two loop halves of the gradiometer $2f_\varepsilon\Phi_0 =(f_1-f_2)\Phi_0\approx0$ the system has a double well potential [Fig.~1(b)]. Here $\Phi_0$ is the magnetic flux quantum $h/(2 e)$ and $f_{\varepsilon,1}$, $f_{\varepsilon,2}$ are the fluxes in units of $\Phi_0$. The ground states in each well of the potential are persistent current states $|L\rangle$ and $|R\rangle$ characterized by the currents $\pm I_p$ carried by the junctions, generating the before mentioned $+/-$ magnetic moments [Fig.~\ref{scheme}(a)]. The energy eigenstates of the qubit are linear superpositions of $|L\rangle$ and $|R\rangle$ [Fig.~\ref{scheme}(c)], following the Hamiltonian
\begin{equation}
\label{H}
H=-\frac{h}{2}\left(\varepsilon(f_\varepsilon,f_\alpha)\sigma_z+\Delta(f_\alpha)\sigma_x\right),
\end{equation}
where $h\varepsilon=2I_p f_\varepsilon\Phi_0$ is the magnetic energy bias and $\sigma_{x,z}$ are Pauli matrices. The critical currents of the four junctions are designed such that the two parallel junctions each have half the value of the critical current $I_0\simeq700$~nA of the other two junctions. Applying ${\it in situ}$ flux $f_\alpha\Phi_0$ to the parallel junctions sets their total effective critical current to $I_0\cos(\pi f_\alpha)$, in this way allowing $f_\alpha$ to control $E_B$ and $\Delta$~\cite{floor}.
Note that, while $\Delta(f_\alpha)$ depends strongly on $f_\alpha$, the persistent current magnitude $I_p$ only shows a weak dependence. Qubit excitation is obtained by the magnetic field generated by current in the symmetrically-split $I_\varepsilon$  line, acting on the qubit flux $f_\varepsilon\Phi_0$. Similarly, the line $I_\alpha$  together with the homogeneous field $B$ generated by an external coil, sets $f_\alpha\Phi_0$  and changes $\Delta$. The geometrical symmetry leads to independent control of $\varepsilon$ and $\Delta$. The qubit states are detected with a DC-SQUID which is coupled to the qubit by a shared wire with  a mutual qubit-SQUID inductance $M\simeq6$~pH.
\begin{figure}[t!]
\includegraphics[width=85mm]{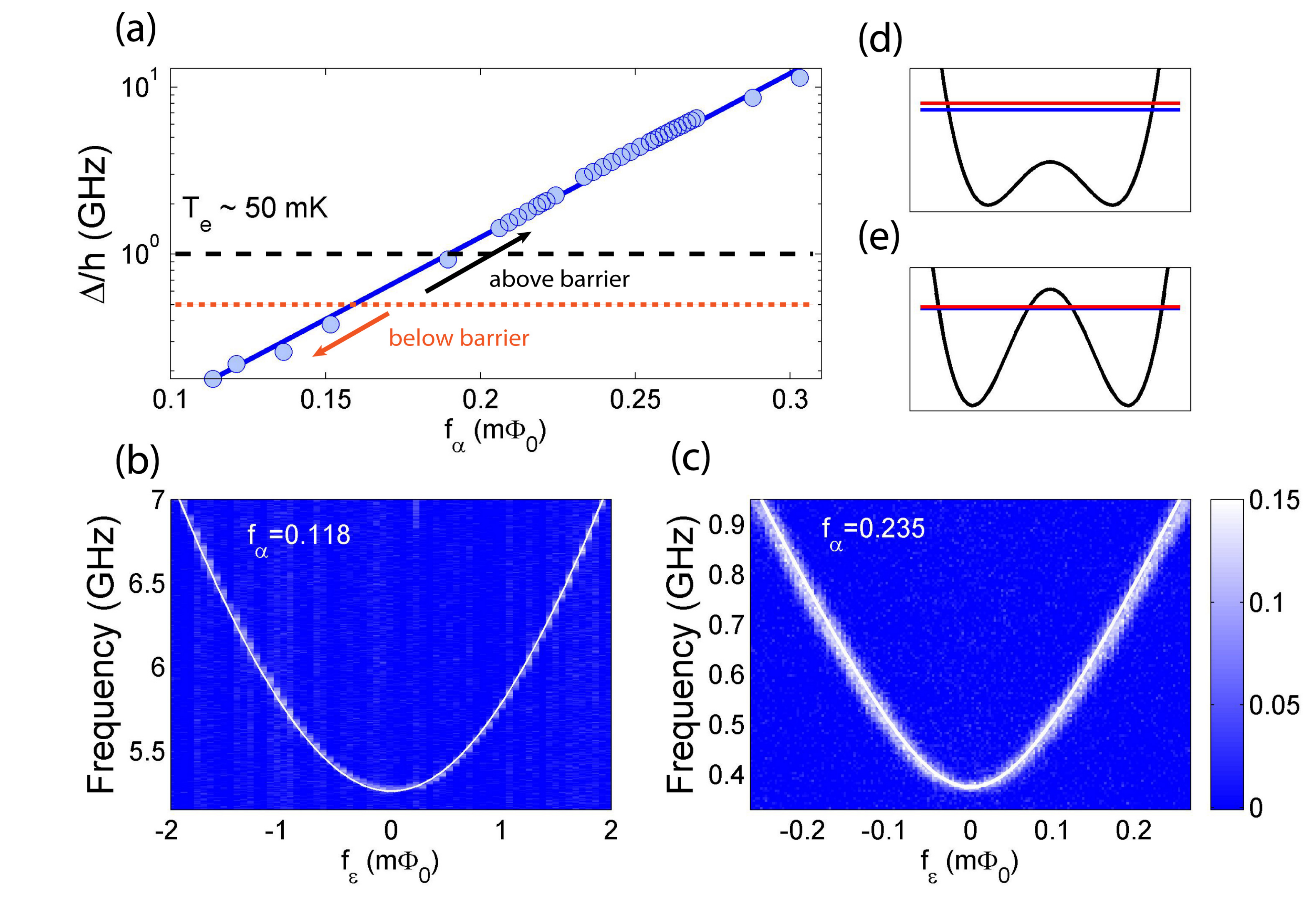}
\caption{\label{spectra} Qubit properties.(a). Gap vs. magnetic frustration $f_\alpha$. The blue line is a guide to the eye with exponential dependence of $\Delta$ on $f_\alpha$. The black dashed line indicates the expected thermal noise level corresponding to $50$ mK; the red dotted line shows the border between formally quantum tunneling (qubit energies are below the tunnel barrier) and quantum scattering (energies are above the barrier) regimes. (b) Spectrum of the qubit in the regular regime for $\Delta = 5.2$ GHz. The white line is a fit with $I_p = 544$ nA.   (c) Spectrum of the qubit in the deep tunnelling regime for $\Delta = 375$ MHz.  The white line shows a fit with $I_p = 344$ nA. (d,e,) numerical simulations of the double well potential and the two lowest states corresponding to the spectra (b) and (c). The color scale represents the SQUID switching probability minus 0.5.}
\end{figure}

Figure~\ref{spectra}(a) shows the gap of the qubit for different $f_\alpha$ and deduced from spectroscopy performed with the following protocol.  First we set $\Delta$ with the field $B$ and apply a DC current $I_{\varepsilon, dc}$ to tune the qubit frequency to $\nu_{\rm qb}\equiv(\Delta^2+\varepsilon^2)^{1/2} \sim 9$~GHz. In the second step we apply a square current pulse $I_{\varepsilon}$, shifting the qubit frequency, combined with a microwave excitation. Next, the qubit  is returned to $\nu_{\rm qb}=9$~GHz and a short bias current pulse $I_b$ is applied to the SQUID detector to measure the qubit state. The relative populations of the qubit ground and excited states determine the expectation value of the persistent current, generating an additional magnetic flux in the SQUID, resulting in a change of the SQUID switching probability (color scale)~\cite{Patrice}. The measurement sequence is repeated a few thousand times to improve signal statistics.

The sequence starts at $\nu_{\rm qb} \sim 9$~GHz, far above the effective noise temperature $T_{\rm e}\sim 50-100$~mK ($\sim$~1-2~GHz) and the cryostat base temperature $T_{\rm b}\cong 20$~mK. After waiting for  a long enough time the qubit relaxes to its ground state. During later operations the qubit splitting can be reduced to values below $T_{\rm e}$ or even $T_{\rm b}$. With the ability of producing fast energy shifts we achieved full coherent control of the qubit quantum state even for very small energy splittings for a duration limited by the relaxation time $T_1$. Coherent transitions below the thermal energy have been realized previously in  superconducting qubits only with active microwave pulses~\cite{orlando,Fluxonium} similar to laser cooling used in atomic systems \cite{Bloch}.

In Fig.~\ref{spectra}(a) one can see that the gap covers nearly two decades, ranging  from 150 MHz to 12 GHz, allowing us to study coherent transitions even in a regime of very small gaps. Over the same $f_\alpha$ range $I_p$ varies from $600$~nA to $150$~nA. Figs.~\ref{spectra}(b,c) show spectra for two representative cases. For the regular flux qubit gap range ($\Delta\sim2-10$~GHz) our numerical simulations show that the qubit ground state level lies above the barrier for the double well potential [Figs.~\ref{spectra}(b,d)]. Only when the gap $\Delta$ drops below $500$~MHz the qubit levels fall below the barrier and the transitions between the wells become classically forbidden [Figs.~\ref{spectra}(c,e)].
Note that $\Delta$ closely follows an exponential dependence on $f_\alpha$ over the full range of $\Delta$, a feature exclusively associated with quantum tunneling.
\begin{figure}[h!]
\includegraphics[width=75mm]{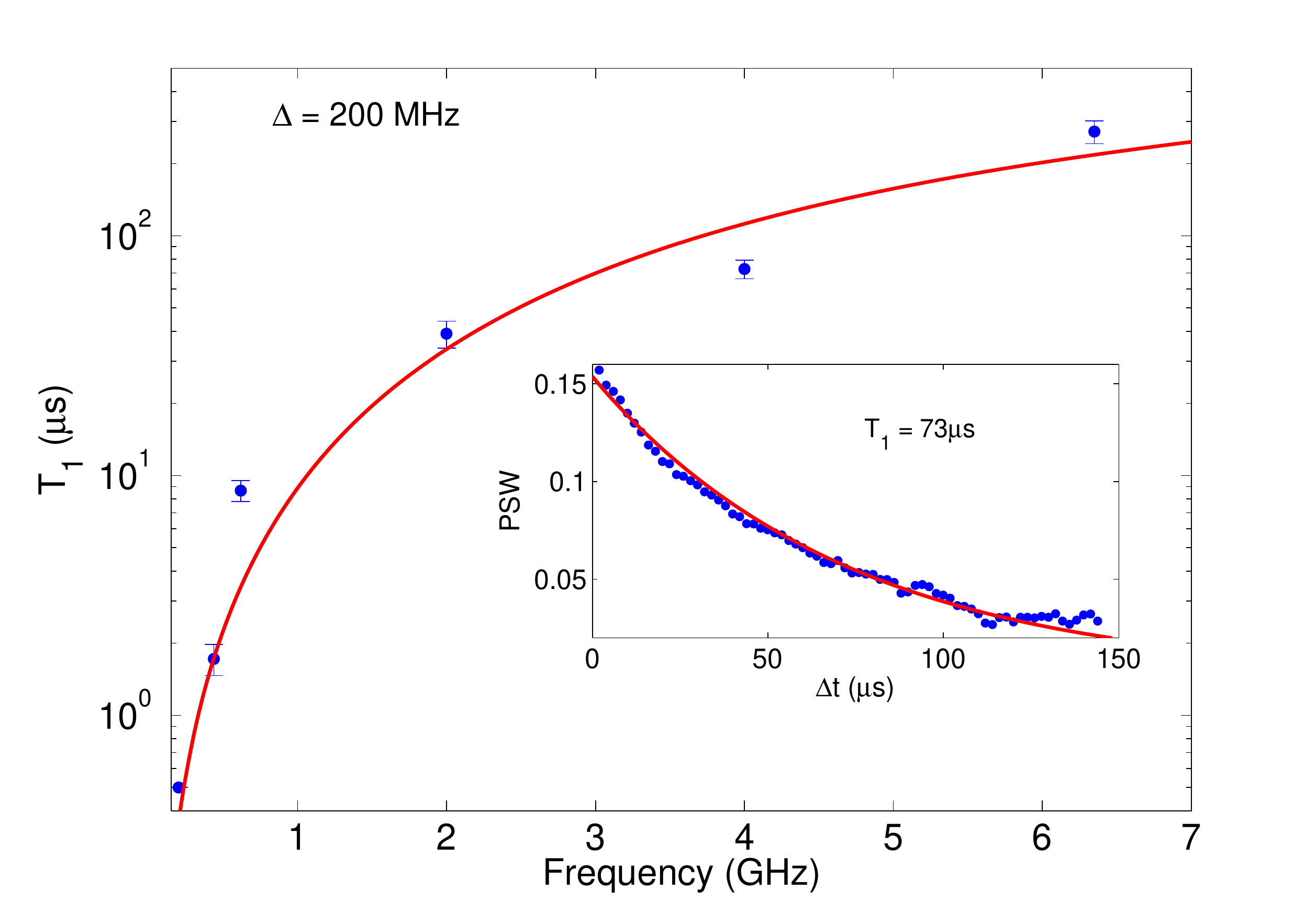}
\caption{\label{T1} Energy relaxation time $T_1$ vs. qubit frequency $\nu_{\rm qb}$ at $\Delta = 200$ MHz. The blue dots show $T_1$ obtained by fitting to the experimental traces measured at each qubit frequency. The error bars are the confidence intervals of the fits. The red line indicates the expected relative dependence $T_1^{-1}\propto\cos^2\theta\nu_{\rm qb}\coth\left(h \nu_{\rm qb}/2k_BT_e\right)$ with $\tan\theta\equiv\Delta /\varepsilon$ with the absolute magnitude being a fit parameter. The noise was assumed to be Ohmic with  $T_e = 50$~mK. The inset shows the actual data trace for $\nu_{\rm qb}=4$ GHz with $T_1 = 73\;\mu$s (the red line is fit to the measurement data). $PSW$ is the switching probability of the SQUID minus 0.5.}
\end{figure}

In order to demonstrate the emergence of the classical opaqueness of the barrier and the macroscopic nature of the persistent current states we measured the relaxation time for a small gap $\Delta = 200$~MHz as a function of the qubit frequency $\nu_{\rm qb}$~[Fig.~\ref{T1}]. From (\ref{H}) it follows that $|0\rangle (|1\rangle)\propto\left[1+(-) \cos \theta\right]|L\rangle+\sin \theta |R\rangle$, where $\tan\theta\equiv\Delta /\varepsilon$. Thus starting from $\nu_{\rm qb}=\Delta$ the energy eigenstates are gradually transformed from (anti)symmetric superpositions of $|L\rangle$ and $|R\rangle$ states to being almost purely $|L\rangle$ and $|R\rangle$ at $\nu_{\rm qb}=6~{\rm GHz}\gg\Delta$. The measurement shows nearly three orders of magnitude increase in lifetime of the excited state for the localized persistent current state $|R\rangle$ compared to the delocalized superposition $(|L\rangle-|R\rangle)/\sqrt{2}$, reaching hundreds of $\mu$s. These high values of $T_1$ demonstrate the extreme robustness of the persistent current states.
\begin{figure}[t!]
\includegraphics[width=90mm]{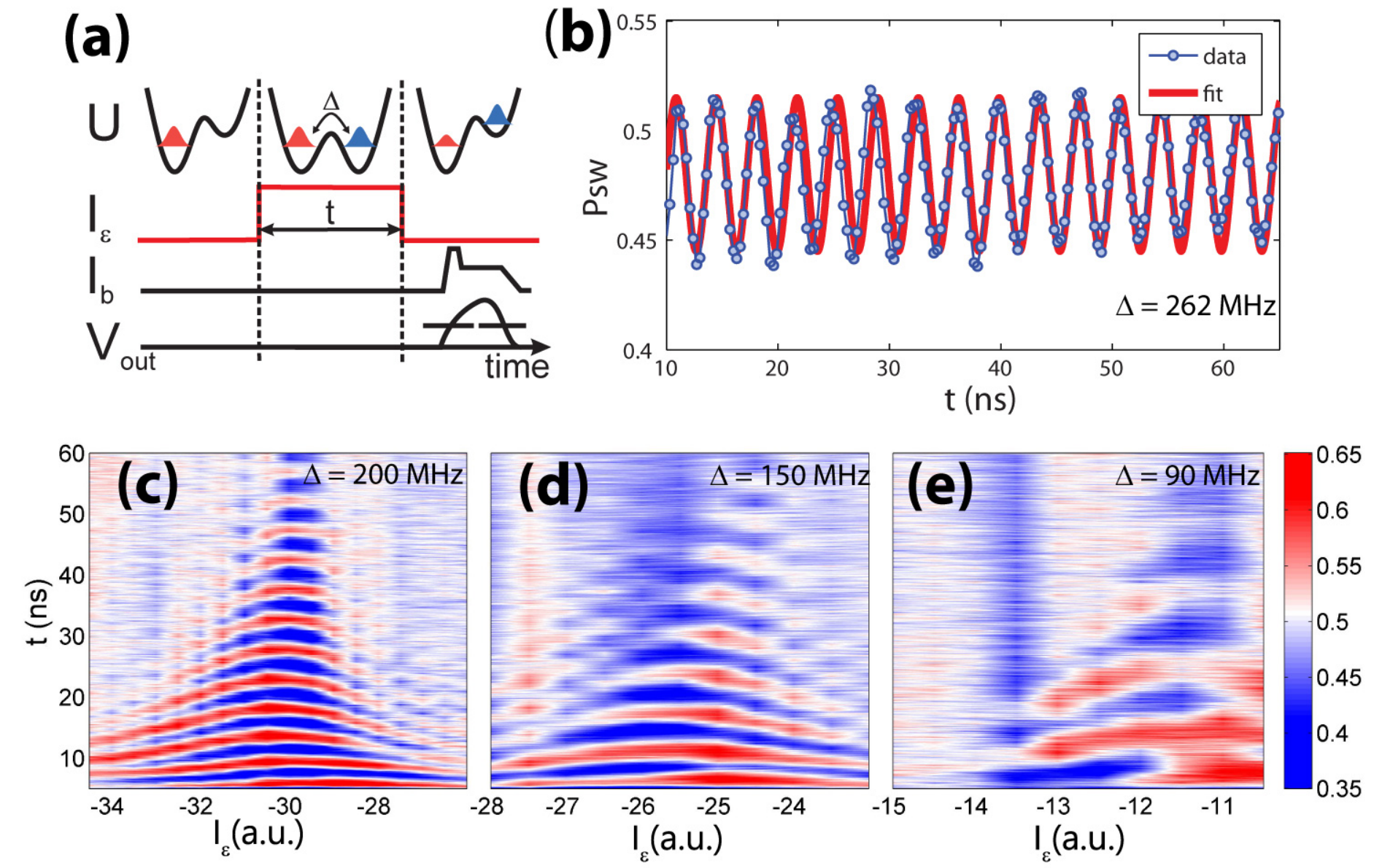}
\caption{\label{mqc} Macroscopic quantum coherence of the persistent current states. (a) Measurement protocol: preparation of the qubit in $|L\rangle$ at $\nu_{\rm qb}=7$~GHz by strongly tilting the double-well potential; a fast shift to the symmetry point (symmetric double-well); free evolution for time $t$ and a fast shift back to $\nu_{\rm qb}$ followed by the SQUID measurement pulse.
(b) Time-resolved measurement of the tunnelling of the persistent current states for $\Delta=262$~MHz (blue) and fit to $\cos(\Delta t)$(red).
(c-e) The colors indicate the switching probability of the SQUID. The horizontal scale represents the amplitude of the current pulse $I_\varepsilon$ sweeping the qubit through the symmetry point where the macroscopic quantum coherence oscillations are clearly observed  for $\Delta=200$~MHz (c) and  $\Delta=150$~MHz (d), Residual oscillations are seen even for $\Delta=90$~MHz (e) representing the boundary between quantum and classical state of the system.
}
\end{figure}

We used the experimental sequence shown in Fig.~\ref{mqc}(a) for time-resolved detection of macroscopic quantum coherence. We start by tuning $\Delta$ below $300$~MHz with the magnetic field $B$ to enter the deep tunnelling regime. Using $I_{\varepsilon,dc}$ we also tilt the double-well potential, preparing the qubit in its ground state $|L\rangle$ with $\nu_{\rm qb}=7$~GHz. Subsequently, the double-well is made symmetric by means of a fast $I_\varepsilon$ pulse; in $0.3$~ns the qubit is taken to its symmetry point. As the qubit energy changes fast relative to the tunnelling amplitude $\Delta$, this transfer is non-adiabatic thus preserving the initial state occupation. These operations lead to the situation described in Fig.~\ref{double well}, where the system is prepared in one of the wells of the symmetric double well potential with a classically impenetrable barrier. The qubit is kept here for a time $t$, then returned fast to the 7~GHz level and finally read out to complete the `real-time' experiment~\cite{Leggett review}. The resulting macroscopic quantum oscillations are shown in Fig.~\ref{mqc}(b).

In our final measurement we monitored the tunnelling  of the macroscopic persistent current states  while successively increasing the tunnel barrier [Fig.~\ref{mqc}(c-e)]. Each experiment is performed by sweeping both the length and the amplitude of the $I_\varepsilon$ pulse.
As the barrier is raised the oscillations become slower, as expected from the corresponding decrease in the tunnel coupling $\Delta$. The oscillation decay is caused by dephasing of the system and is characterized by the dephasing time $T_2$. The longest decay time is achieved around the symmetry point  where the influence of the low frequency flux noise in $f_\varepsilon$ is suppressed. With higher barriers the sensitivity to low frequency flux noise increases and the phase coherence decays faster. The slowest oscillations are observed for $\Delta=90$~MHz (i.e. an energy splitting equivalent to only 4~mK). Here the oscillation period approaches the dephasing time, thus showing the border between quantum and classical regime. A further increase of the barrier leads to a total destruction of the quantum phase between the persistent current states and the system is no-longer regarded as quantum. It is interesting to note that, for small gaps, sensitivity to $f_\alpha$ noise is strongly suppressed, and so the environment automatically chooses the basis of the macroscopic current states for dephasing.

Our measurements shows how the flux qubit can be gradually tuned from the quantum to the classical regime. With the increase of the tunnel barrier the `quantumness' of the system manifested in coherent tunnelling is gradually lost. At the same time the lifetime of the persistent current states dramatically increases, which is naturally associated with macroscopic classical systems or classical bits. Also, over a large range of parameters the quantum and macroscopic properties are shown to coexist. Our experiment demonstrates the potential of fabricated quantum objects, where knobs are available to tune parameters {\it in situ}, for fundamental research as well as for applications.

We thank R.~N.~Schouten for technical support. This work was supported by the Dutch NanoNed program, the Dutch Organization for Fundamental Research (FOM), and the EU projects EuroSQIP, CORNER and SOLID.

\end{document}